\documentclass[conference]{IEEEtran}
\pdfoutput=1
\ifCLASSINFOpdf
\else
\fi
\IEEEoverridecommandlockouts
\usepackage{cite}
\usepackage{amsmath,amssymb,amsfonts}
\usepackage{algorithmic}
\usepackage{graphicx}
\usepackage[caption=false]{subfig}
\usepackage{textcomp}
\usepackage{xcolor}
\usepackage{color, soul}
\usepackage{enumitem}
\usepackage{import}
\usepackage[ruled,vlined]{algorithm2e}
\usepackage{setspace}
\usepackage{comment}
\usepackage{hyperref}
\usepackage{tikz}
\usepackage{tabularx}
\usepackage{epsf}
\usepackage{multirow}
\usepackage{tablefootnote}

\def\BibTeX{{\rm B\kern-.05em{\sc i\kern-.025em b}\kern-.08em
    T\kern-.1667em\lower.7ex\hbox{E}\kern-.125emX}}

\begin{document}

\title{Security Assessment and Impact Analysis of Cyberattacks in Integrated T\&D Power Systems}

\IEEEaftertitletext{}
\author{\IEEEauthorblockN{\textbf{Ioannis Zografopoulos}\IEEEauthorrefmark{1}
\thanks{
This work was supported in part by the U.S. Department of Energy’s Office of Energy Efficiency and Renewable Energy (EERE) under the Solar Energy Technology Office (SETO) Award Number DE-EE0008768.}, \textbf{Charalambos Konstantinou}\IEEEauthorrefmark{1},\\ \textbf{Nektarios Georgios Tsoutsos}\IEEEauthorrefmark{2}, \textbf{Dan Zhu}\IEEEauthorrefmark{3}, \textbf{Robert Broadwater}\IEEEauthorrefmark{3}}
\IEEEauthorblockA{\IEEEauthorrefmark{1}FAMU-FSU College of Engineering, 
Center for Advanced Power Systems, Florida State University}
\IEEEauthorblockA{\IEEEauthorrefmark{2}Department of Electrical and Computer Engineering, University of Delaware}
\IEEEauthorblockA{\IEEEauthorrefmark{3}Electrical Distribution Design} Email:\{izografopoulos, ckonstantinou\}@fsu.edu \\ tsoutsos@udel.edu, \{dan.zhu, robert.broadwater\}@nisc.coop\vspace{-3mm}}

\IEEEaftertitletext{}
\maketitle

\begin{abstract}
In this paper, we examine the impact of cyberattacks in an integrated transmission and distribution (T\&D)  power grid model with distributed energy resource (DER) integration. We adopt the OCTAVE Allegro methodology to identify critical system assets, enumerate potential threats, analyze, and prioritize risks for threat scenarios. Based on the analysis, attack strategies and exploitation scenarios are identified which could lead to system compromise. Specifically, we investigate the impact of data integrity attacks in inverted-based solar PV controllers, control signal blocking attacks in protective switches and breakers, and coordinated monitoring and switching time-delay attacks.  
\end{abstract}


\section{Introduction}
The power grid is the largest machine ever built. 
Electrical grids started to surface in the late 19th century providing energy to consumers, but a lot has changed since then. Nowadays, modern grid deployments enable flexible control over power generation to cover the current demand. In addition, grid modernization efforts aim to upgrade the legacy infrastructure and improve power generation and dispatch leveraging information and communication technologies (ICT) as well as renewable and distributed energy resources (DERs)\cite{muyeen2017communication, konstantinou2021secure}. DERs, being small generation or storage systems, such as rooftop solar, and battery storage, apart from their much lower deployment and operation overheads, can be placed close to distribution-level consumers allowing on-site power generation and consumption, minimization of delivery costs, and increased grid resilience due to the generation redundancy.

    \begin{figure*}[t!]
        \centering
        \includegraphics[width=0.8\textwidth]{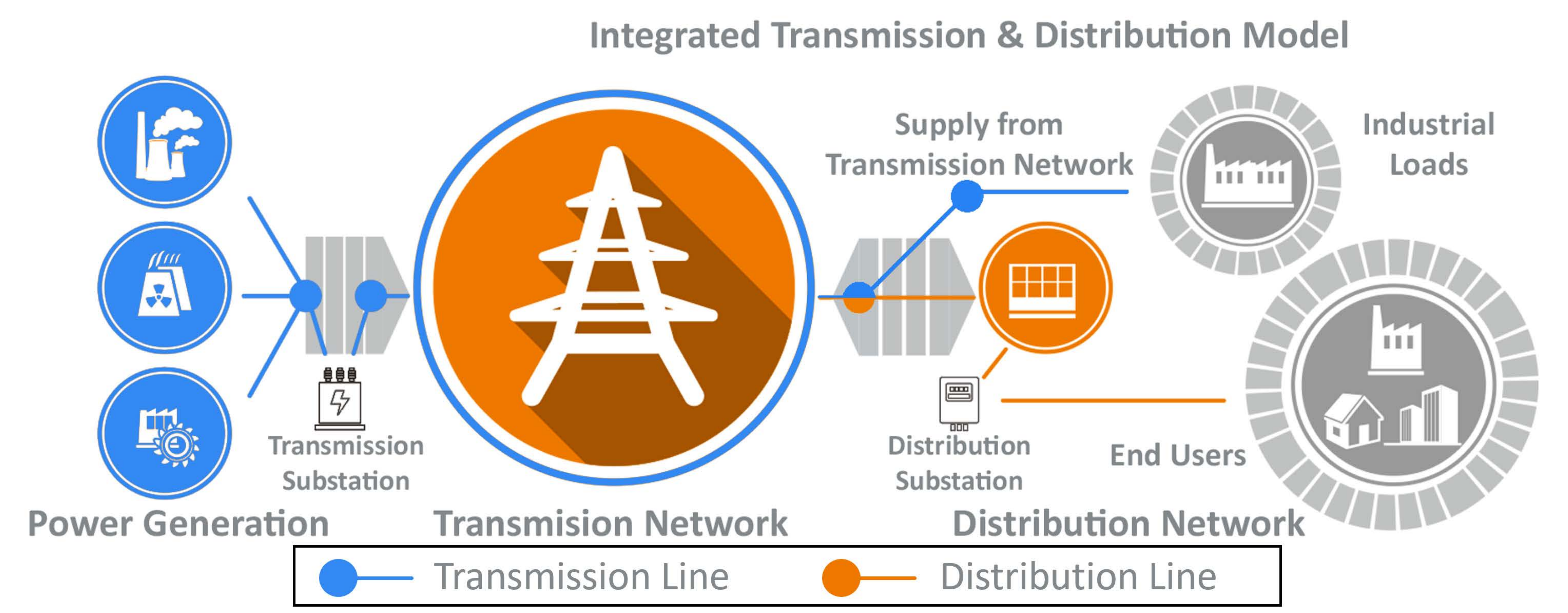}
        
        \caption{Integrated transmission and distribution (T\&D) model.}
        \vspace{-2mm}
        \label{fig:integrModel}
        
    \end{figure*}

The increasing penetration of DERs and the ICT integration emphasizes the need for understanding the interdependency of the interactions between distribution and transmission systems. In the past, power system studies were conducted by modeling and simulating the transmission and distribution (T\&D) systems independently. Recent works, however, utilize integrated T\&D models demonstrating that this approach can capture grid synergies with high fidelity \cite{jain2016three, tbaileh2017graph, bhatti2020analyzing}.  
Fig. \ref{fig:integrModel} illustrates the top level architecture of an integrated T\&D system. Such models are crucial in investigating the effects of distribution systems' anomalous operation to the transmission systems, and vice versa, as well as the impact of cyberattacks holistically.

The proliferation of smart meters and smart inverters increases the threat surface and exposes the power grid to greater risk of cyberattacks \cite{kuruvila2020hardware}. Thus, threat modeling and risk assessment are important tools to identify and evaluate potential threats, as well as prioritize the corresponding risks to the power system and administer mitigation strategies. In this work, we employ the \emph{OCTAVE Allegro} methodology to identify critical system assets and enumerate potential threats. We perform a comprehensive analysis for threat scenarios and prioritize attack risks based on their expected outcomes on system operation. Moreover, our cybersecurity analysis demonstrates the impact evaluation of the identified threat scenarios. Specifically, we perform an impact analysis study for three main attack categories on an actual integrated T\&D model and dataset.

The roadmap of the paper is as follows. Section \ref{s:threatModel} presents the background and risk assessment method. Section \ref{s:attackClass} describes the attack classes, adversary objectives, and potential attack outcomes. Section \ref{s:results} presents the simulation setup and experimental results, while Section \ref{s:conclusions} concludes the paper.

\vspace{-1mm}
\section{Threat Modeling and Risk Assessment} \label{s:threatModel}

Over the past years, power systems have experienced drastic transformations to address the growth in energy demand and enhance power quality and energy efficiency. The shift to the smart grid involves, among others, the inclusion of smart inverters, intelligent electronic devices (IEDs), and advanced metering infrastructure (AMI). 
Embedded device controllers are used to support the communication and control functions of inverters \cite{qi2016cybersecurity}. Additionally, grid assets and their operation mechanisms (e.g., switches, breakers) are often controlled using IEDs \cite{konstantinou2015impact}. Furthermore, AMI, such as smart meters and monitor points (MPs), enables better situational awareness and helps detect anomalous system behavior and cyberattack intrusions. 
The inclusion of the aforementioned components within T\&D systems, however, increases the threat surface. Vulnerabilities of such units can be ported to the power grid \cite{glenn2016cyber}, while insecure control networks and protocol implementations further exacerbate the problem \cite{johnson2017roadmap, zografopoulos2020derauth, zografopoulos2020harness}. 

We refer to mission-critical system assets that can jeopardize grid operations if compromised by malicious actors as {\it crown-jewels} \cite{crownmitre}. Notably, these devices include grid inverters, utility-to-device communication channels, physical interfaces, substation circuit breakers/reclosers, and controllers. Gaining access to any of these assets can enable an adversary to manipulate the generated or stored energy, cause switch disconnections altering the system topology, false trips, feeder overloadings, voltage-frequency violations, damage protection equipment, or inflict system instabilities \cite{johnson2017roadmap, qi2016cybersecurity, 9308900}. In addition, the grid communication infrastructure and industrial protocols could be targeted by adversaries to mount their attacks. For example, attacks targeting DNP3 communications could exploit vendor implementation issues of the protocol, protocol specification vulnerabilities, and/or vulnerabilities in the supporting communication infrastructure. According to the Electric Power Research Institute (EPRI), more than 75\% of North American electric utilities use the DNP3 protocol for industrial control applications and supervisory control and data acquisition (SCADA) systems \cite{jin2011event}.

{A multitude of standards, recommendations, technical notes, and best practices exist to support and protect the operation of critical infrastructure such as power systems \cite{alcaraz2015critical}. Typical examples include NIST SP 800-82 and IEC 61850, which provide guidelines to maintain system  reliability, interoperability, and advanced protection and secure control for industrial control systems (ICS), SCADA, distributed control systems, and/or other industrial applications \cite{NISTSP}. 
The NISTIR 7628 framework provides guidelines enhancing smart grid cybersecurity and promotes the deployment of 
security schemes that factor the unique characteristics and vulnerabilities (e.g.,  
distributed interconnected nature, communications, etc.) of diverse smart grid deployments \cite{NISTIR}. Apart from the aforementioned standards, other security frameworks include NERC critical infrastructure protection (CIP), ISA99, IEEE 1402 (for physical security), etc. Although the recommendations discussed in such standards can contribute towards effectively protecting critical infrastructure and power systems, they serve as security recommendations with limited enforcing capabilities, therefore, complimentary security assessments should also be performed.}

{Security assessment should be an integral part of every cyberphysical system's security analysis, and in the light of mission-critical power systems, its importance is stressed even more. Security assessment provides a comprehensive overview of the system assets and their underlined interconnections,  and certifies that they satisfy the requisite security specifications denoted by the current standards. Additionally, frequent assessments enable security analysts to identify potential vulnerable, outdated, or malfunctioning system components and update or replace them. 
Thus, the overall infrastructure remains secure and reliable, by auditing the system components and enforcing 
security policies.}

\subsection{Security Assessment with OCTAVE Allegro}\label{ss:octave}

The power grid security assessment in our work is performed using the OCTAVE Allegro risk assessment methodology \cite{zografopoulos2021cyberphysical}. 
The first step of the analysis entails: \textit{(i)} identification of critical system assets, \textit{(ii)} identification of security requirements, and \textit{(iii)} identification of security threats to the critical assets. The second step focuses on: \textit{(i)} identifying the criteria for impact evaluation when a threat is realized, \textit{(ii)} defining the priority/importance of the identified impact evaluation areas, and \textit{(iii)} calculating the relative risk to each critical asset based on the probability and impact of the applicable threats. The third step defines strategies to manage the identified risks.

A threat refers to a situation or scenario in which an entity (e.g., a threat actor) or natural occurrence could cause an undesirable outcome. During the first step of OCTAVE Allegro, the security threats applicable to the critical assets are identified. Each threat is associated, and later analyzed according to its corresponding parameters: actor, affected asset, outcome, motive, and access. Next, the threat scenarios are defined to show how a system asset is compromised if an actor, who has a motive and an access method, causes an undesired outcome to the target asset. In essence, the devised threat scenarios are useful for articulating the existing risks to critical assets. 

\vspace{-4mm}
{\subsection{Security Assessment Methodology} }
\vspace{-1mm}

In our risk analysis, we focus on three core components of the grid infrastructure. We assume that the threat actor is a malicious adversary with deliberate motives to compromise the system. Specifically, we investigate scenarios involving DER and inverter control, SCADA controlled devices (e.g., switches, reclosers, breakers), and MPs (e.g., smart meters). In Table \ref{t:threatScenarios}, we demonstrate the three aforementioned critical assets alongside the respective threat scenarios, risk probabilities, severity, and comprehensive risk scores. The risk probability for each scenario is determined by the security analysis team and reflects how plausible it is for such an attack to occur based on the asset's location in the system, cyberphysical security perimeter, etc. 
{Risk probability is qualitatively assessed, receiving scores Low ($1$), Medium ($2$), or High ($3$).}

\begin{table}
\small
\setlength{\tabcolsep}{1.2pt}
\centering
\caption{Critical Asset Threat Scenarios.}

\label{t:threatScenarios}

\renewcommand{\tabularxcolumn}[1]{m{#1}}

    \begin{tabularx}{\linewidth} { 
      || >{\hsize=.35\hsize\linewidth=\hsize\centering\arraybackslash}X 
      | >{\hsize=.7\hsize\linewidth=\hsize\centering\arraybackslash}X
      | >{\hsize=.45\hsize\linewidth=\hsize\centering\arraybackslash}X
      | >{\hsize=.3\hsize\linewidth=\hsize\centering\arraybackslash}X
      | >{\hsize=.2\hsize\linewidth=\hsize\centering\arraybackslash}X || }
      
     \hline
     \hline
     \centering\arraybackslash\textbf{Affected Asset*} & {\textbf{Outcome*}} & \textbf{Risk Probability}  & \textbf{Severity} &\textbf{Risk Score} \\ 
     \hline \hline
     {\textbf{Solar Inverters}} & { Transient Voltage \& Frequency Instability}  & {Medium} & {27} & {54} \\
    \hline
    {\textbf{SCADA Devices}} & {Anomalous Grid Sectionalization \& Electricity Loss}  & {Low} & {31} & {31}  \\
    \hline
    {\textbf{Monitor Points}} & {Loss of Situational Awareness \& Erroneous Control}  & {Medium} & {15} & {30}  \\
    \hline
    \hline 
    \end{tabularx}
    \raggedright\small{*Assumptions:~Threat actor = attacker, motive = deliberate, and the access = via technical means (i.e., without physical access) }
    \vspace{-1mm}
\end{table}

{Severity reflects the effect of the scenario to the grid operation. To calculate the severity index of each threat scenario, i.e.,  the effect on the grid operation, the attack impact areas and their corresponding impact scores need to be delineated. The number and type of impact areas are determined by the security team based on the system under investigation and mission criticality. After the impact areas are identified, they are prioritized depending on the asset's objective within the system, with scores from $1$ to $n$, with $n$ being the most significant. For our study, we have selected $5$ impact areas (i.e., $n=5$) which are the following: ``safety and health'', ``financial'', ``productivity'', ``reputation'', ``fines and legal penalties'' and their respective priorities are provided (in descending order) in Table \ref{table:priority}. The threat impact score on each of the areas is then qualitatively assessed, receiving scores equal to Low ($1$), Medium ($2$), or High ($3$).}

\begin{table}[t]
\setlength{\tabcolsep}{1.2pt}
\centering
\caption{Risk Assessment Impact Area Priorities. }
\label{table:priority}
\begin{tabularx}{0.85\linewidth} { 
      || >{\hsize=1.5\hsize\linewidth=\hsize\centering\arraybackslash}X 
      | >{\hsize=.5\hsize\linewidth=\hsize\centering\arraybackslash}X || }
      
     \hline
     \hline
     \textbf{Impact Area} & {\textbf{Priority}} \\ 
     \hline \hline
     {Safety and health} & {5} \\
     \hline
     {Financial} & {4} \\
     \hline
     {Productivity} & {3} \\
     \hline
     {Reputation} & {2} \\
     \hline
     {Fines and legal penalties} & {1} \\
     \hline
     \hline
\end{tabularx}
\vspace{-2mm}
\end{table}

{After the impact areas are defined, prioritized, and associated with their respective impacts, the severity for each threat scenario is evaluated. Severity is calculated using Eq. \eqref{Eq:severity}.}
\begin{equation}
\begin{array}{l}
    Severity = \sum_{1}^{n} (Impact \ Area) \times (Impact \ Score)
\label{Eq:severity}
\end{array}
\end{equation}

\noindent  
{The comprehensive risk score for each affected asset is then calculated using Eq. \eqref{Eq:risk}
and can be leveraged for one-to-one comparisons between assets aiding the prioritization of those with the highest relative risks.}
\begin{equation}
\begin{array}{l}
    Risk \ Score = Risk \ Probability \times Severity
\label{Eq:risk}
\end{array}
\end{equation}

{For the solar inverter threat scenario described in Table \ref{t:threatScenarios}, the calculation of the comprehensive risk score is as follows. The risk probability for this threat scenario is set to Medium ($2$), since the wide adoption of inverter-based resources and their remote communication and control functions -- relying on ICT -- introduce more potential attack entry points. Given the power systems' mitigation and redundancy mechanisms in place to maintain stability and impede cascading failures or blackouts, the risk probability of this scenario cannot be set to High ($3$). To calculate the severity, impact score values should be associated with the five impact areas that we have selected for our study (Table \ref{table:priority}). Namely, the impact scores for ``safety and health'', ``productivity'', and ``fines and legal penalties'' are set to Low ($1$). On the other hand, the impact scores for the ``financial'' and ``reputation'' impact areas are set to High ($3$) due to the substantial economic and reputation consequences that such threats could introduce for energy providers. Using Eq. \eqref{Eq:severity}, severity is estimated to be $\sum (5*1) + (4*3) + (3*1) + (2*3) + (1*1) = 27$. Similarly,  from Eq. \eqref{Eq:risk}, $Risk \ Score = 2 * 27 = 54$. Following the same methodology, the risk scores for the other two threat scenarios (of Table \ref{t:threatScenarios}) are evaluated.}

\vspace{1mm}
\section{Attack Classes} \label{s:attackClass}
\vspace{-1mm}

\begin{figure}[t!]
    \centering
    \includegraphics[width=\linewidth]{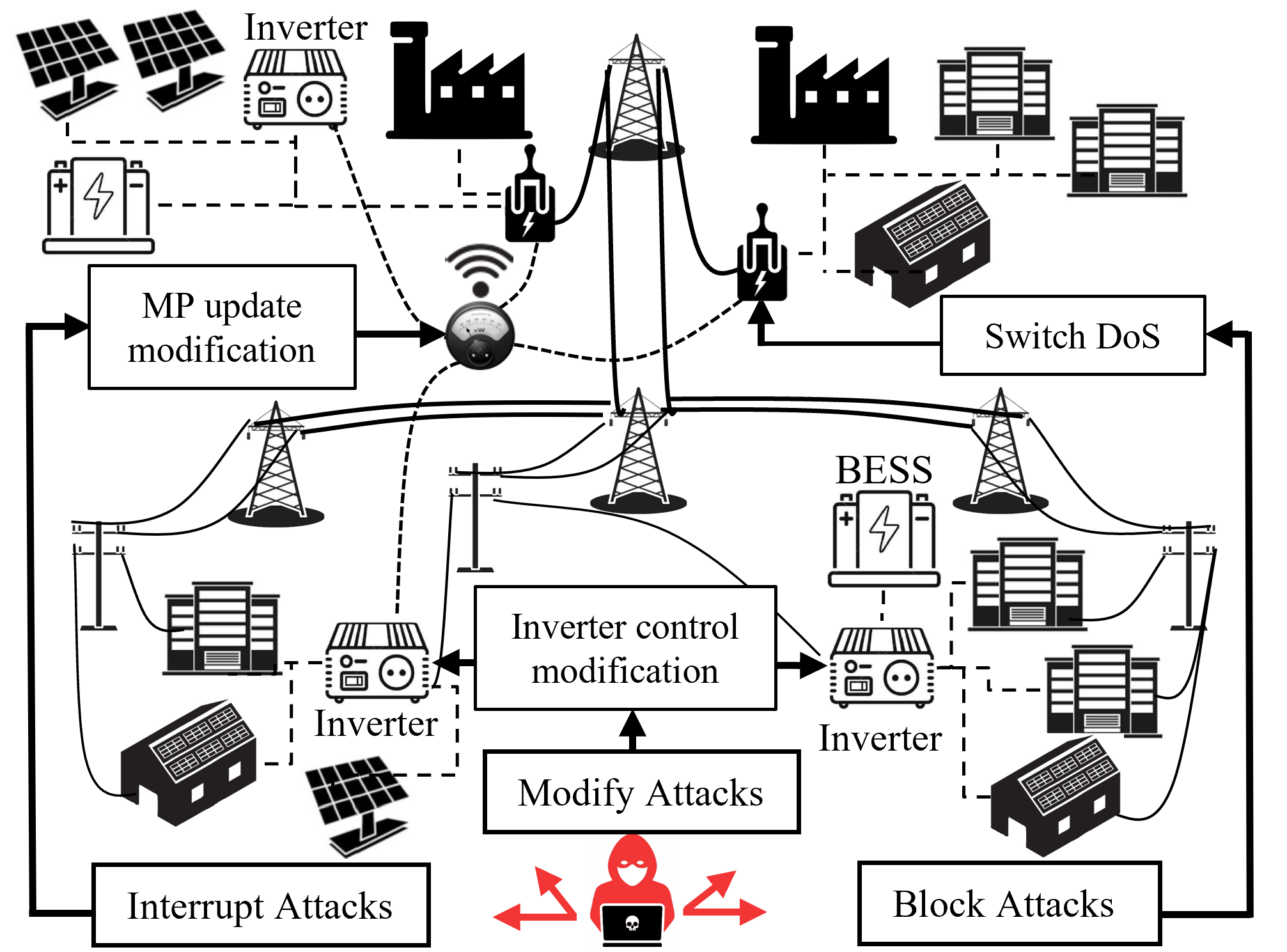}
    
    \caption{Power grid cyberattack scenarios.}
    \label{fig:attacks}
    \vspace{-1mm}
    
\end{figure}

Our analysis focuses on three attack directions, aligned with our threat modeling in Section \ref{s:threatModel} and depicted in Fig. \ref{fig:attacks}: \textit{(i)} data modification attacks, \textit{(ii)} loss/blocking attacks during system-critical operations, and \textit{(iii)} interruption of system-critical operation attacks. 

The first attack category refers to tampering attacks aiming to maliciously modify system data. Data tampering includes attack scenarios in which control commands are manipulated without detection. Such attacks can be launched, for example, via communication channel corruption or exploitation of IED vulnerabilities.  
In the second category, the adversarial objective is to block operational commands in system-critical operations, i.e., commands from authorized entities are blocked when needed. For example, attackers could prevent access by overwhelming a resource with traffic overflowing its network bandwidth. In the third category, interruption of system-critical operations is enabled by delaying commands or data to grid components despite being issued by legitimate operators. This type of time-based attacks could undermine system operations by delaying real-time control signals or measurements \cite{ospina2020demo}.

\subsection{Modification of data: DER integrity attacks}

DERs and supporting inverters serve as ancillary generation sources providing power to the grid. To control inverters and harness the generated power, two main categories of grid functions are implemented: 
\textit{(i)} functions used by operators giving them direct control over the corresponding inverter operation, and \textit{(ii)} autonomous functions which allow inverters to operate independently, making decisions based on their environment (e.g., power demand, generation capacity, connected loads, etc.). Our analysis covers the first type of functions which, among others, include limiting the output power of an inverter, setting active and reactive power limits, changing the power factor of the inverter, as well as controlling volt-var and watt-var operational modes.

Similar to grid operators, malicious users are able to bypass the power system’s security mechanisms, they can modify control commands or issue forged ones, altering the operation of inverters to destabilize the grid. Commands which could be of interest to malicious attackers are: 

\textit{(i)} \emph{Constant power factor (PF) mode}: an inverter is maliciously set to operate at a constant PF, inductive or capacitive, which could potentially create voltage regulation issues, increase system losses, and reduce the electric system power quality.

\textit{(ii)} \emph{Limit active power ($P$) mode}: the amount of $P$ injected by an inverter is maliciously controlled and limited to a setpoint, resulting in curtailing the injected $P$ amount to the grid. 

\textit{(iii)} \emph{Constant reactive power ($Q$) mode}: Similar to the $P$ mode, an inverter can inject or absorb a constant amount of $Q$ defined by a maliciously modified setpoint, causing undervoltage/overvoltage at points of common coupling.

\subsection{Loss/blocking during system-critical operations: switch and breaker control attacks}

Unexpected events can disrupt the steady-state operation of power systems, leading to line overload, frequency deviations, voltage instabilities, or even cascading outages. Such events can either be inadvertent, e.g., component or equipment failure, or intentional in the case of malicious attacks. To deal with the such events, immediate and protective actions should be taken. Typical countermeasures to prevent these undesirable effects and avoid a generalized system collapse involve power generation and dispatch coordination, and system re-configuration via line and bus switching (through  recloser controllers, switches, circuit breakers, etc.) which actively changes the system topology. 

Attacks on switchers and breakers (e.g., by issuing malicious control commands to open/close) could trigger cascaded sequences of events. For example, if attackers gain access to a substation’s ICT network, they could falsify circuit breaker control signals at a targeted IED, causing tripping of the IED-connected breakers. The result of maliciously controlled breakers could violate operational voltage limits and line overload conditions initiating cascading outage events. In essence, the end goal of such attacks is to open or close circuit breakers, change the system topology causing line overloads, and thus lead to serious problems including blackouts, brownouts, equipment failures, and uneconomical system operation.

\subsection{Interruption of system-critical operations: coordinated monitoring and switching attacks}

Protective switches and breakers are designed to handle power network faults (e.g., short-circuits) and sectionalize areas with sufficient response time to minimize fault duration, reroute power flow, and avoid any equipment damage. This involves isolating areas via tripping the breakers and eventually reclosing circuits automatically. This operation attempts to preserve stability and minimize the impact on the rest of the system. Failure to open/close the switch/breaker may initiate chain reactions. In this category of attacks, we delay the control commands to the switching devices despite being issued by legitimate operators. At the same time, MPs due to their sporadic 
and unsynchronized measurements, cannot effectively detect momentary time-delay malicious events. Adversaries can thus exploit the reporting mechanism operation (e.g., infrequent and/or unsynchronized measurements) while remaining hidden. Consequently, grid situational awareness is compromised, monitoring routines cannot detect and promptly initiate mitigation strategies to avoid outage events, and adversaries can stealthily mount their attacks undetected. 

Regardless of the reporting measurement frequency, the possibility for covert attacks persists unless sophisticated 
security schemes are systematically deployed in power systems. Such schemes could collect and verify system event information from different sources (i.e., MPs, SCADA substation, phasor measurement units (PMUs), etc.) ensuring non-repudiation (e.g., triple redundancy checks, consensus algorithms, etc.), impeding adversaries, and enhancing situational awareness \cite{8743447}. Preserving grid stability relies on responding timely to system changes (e.g., faults). Situational awareness is critical for detecting abnormalities, generating automated responses, and mitigating threats. MPs serve as the system's sensors aiding system observability and detecting malicious or anomalous behaviors. Maintaining stable operation relies heavily on retaining visibility of the system states at all times, since adversaries can create transient events that cannot be detected. For instance, short and intermittent malicious events cannot be detected by MPs if their duration is much smaller than the update frequency of the MP (e.g., 15 minutes).

\section{Simulation Setup and Results} \label{s:results}

\begin{figure}[t!]
    \centering
    \includegraphics[width=\linewidth]{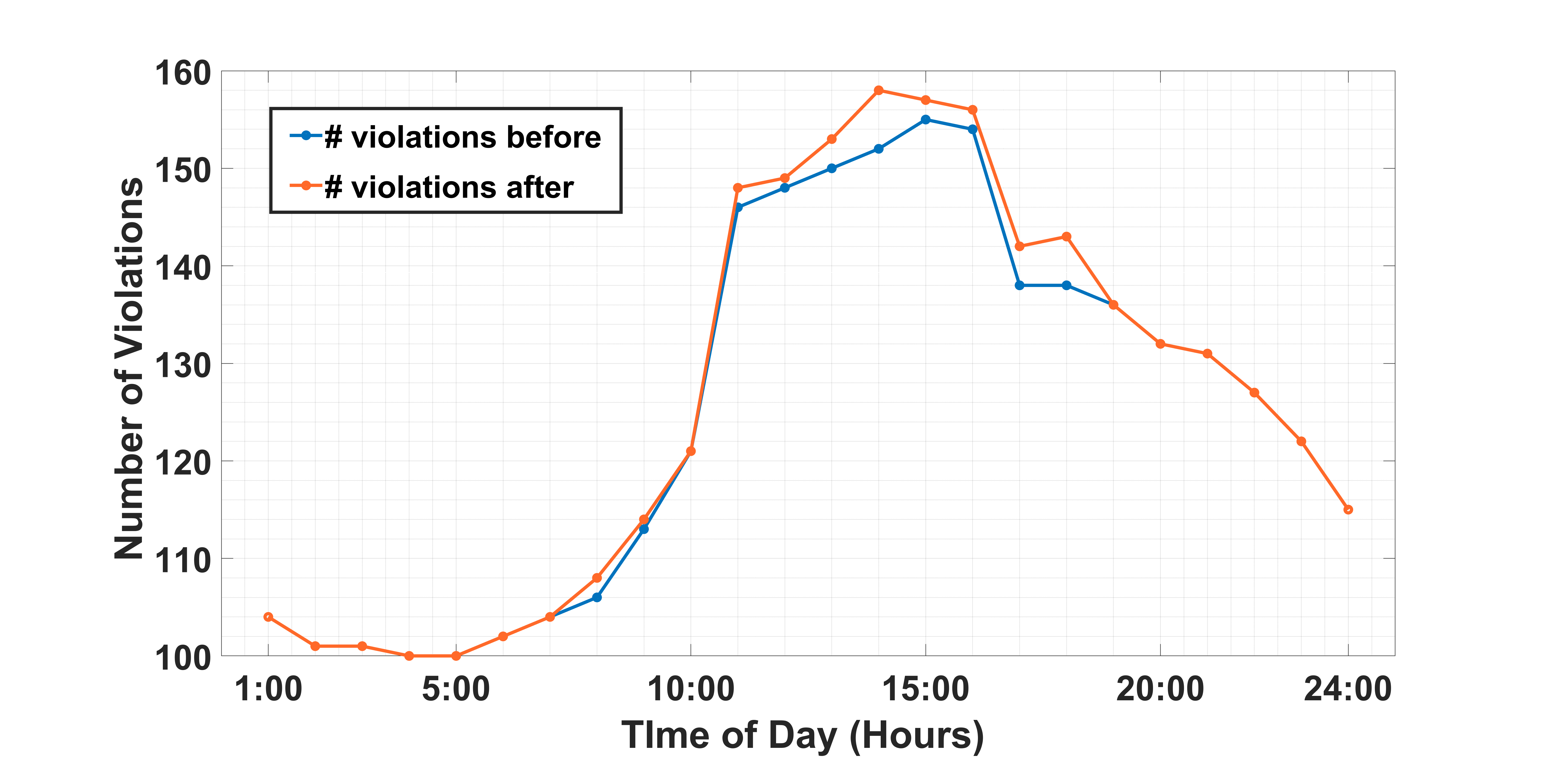}
    
    \caption{Inverter violations before and after compromise.}
    \label{fig:inverter}
    \vspace{-1mm}
    
\end{figure}

Grid devices for power system operation monitoring and controlling can regulate the voltage output and the generated power ($P$, $Q$)  under varying steady-state or transient events. As a result, such devices can control the setpoints of automatic generation control functions, capacitor banks, reactors, load-tap changing transformers, and energy storage and inverter-based resources. Additionally, in many cases, network switching control is utilized by system operators to mitigate component overloading scenarios and other emergencies. The system reconfiguration, i.e., alteration of the system topology via network switching, can involve opening/closing interconnection switches using alternative T\&D lines or splitting busbars to meet power demand \cite{hedman2011review}. These alternative network architectures, although they can mitigate the propagation of adverse effects, they can also lead to uneconomical operation or violation-inducing scenarios \cite{muller1989line}. Thus, in order to evaluate the impact of the cyberattack use cases in the integrated T\&D model, we utilize the number of violations as an indicator before and after the compromise. 

\underline{Violation definition:} With the term violation, we refer to system component behavior exceeding the nominal operational limits and potentially  compromising the stable system operation. For instance, voltage violations are triggered if the voltage at a specific system component surpasses the acceptable range, i.e., higher than $126 V$ (overvoltage), or lower than $114 V$ (undervoltage), for a $120 V$ nominal bus voltage with a 5\% allowed deviation. Similarly, we have current and power violations for components, buses, or lines, if these values exceed the prescribed limits, jeopardizing system operation, equipment performance, and human safety. 

\underline{Simulation setup details:} The simulation, analysis and impact evaluation of the attack classes are performed using the Distributed Engineering Workstation (DEW) simulation software. Additionally, an integrated T\&D model composed of 1834 T\&D load points, 218 solar PV inverters, and 3,000 sectionalizing devices (e.g., cut-out switches, circuit breakers, reclosers, etc.) is employed to highlight the comprehensive impact as well as the interdependency of T\&D networks.

\subsection{Modification of data: DER integrity attacks}
As discussed in Section \ref{s:attackClass}, DERs and inverters can support grid operation providing power either by responding to operator requests (e.g., via issued control commands) or in an autonomous fashion. In this simulation scenario, we assume that an adversary, by compromising the communication infrastructure (i.e., the communication links used by utilities to control DER assets), can modify and inject malicious commands to the deployed inverters. Specifically, the adversary maliciously controls inverters and sets them operating in a purely active ($P$) mode of operation, i.e., the PF is set to $1.0$, while their generation limits ($P$, $Q$) have been decreased inhibiting the inverters to provide power to the grid. To illustrate the grid dependency on inverter-based generated power, we have compiled the voltage, current, and power violation reports corresponding to the aforementioned inverter control modifications. In Fig. \ref{fig:inverter}, we provide a graphical representation of the generated violations throughout the day once the system’s PV inverters get compromised. 
{Notably, during peak working hours the number of violations is higher, compared to early in the morning or late at night when the inverter contribution is expected to be minimal. Thus, determined adversaries could maximize the impact of their attacks targeting DER communications by performing them at hectic periods of the day or during the occurrence of unexpected phenomena (e.g., natural disasters, outages, etc.). On the other hand, issuing setpoint alteration attacks to solar PV inverter during the night, would be a sub-optimal attack tactic since it would incur minimal disturbance on the power system as indicated in Fig. \ref{fig:inverter}. }

\subsection{Loss/blocking during system-critical operations: switch and breaker control attacks}\label{s:switchBlock}

\begin{figure}[t!]
    \centering
    \includegraphics[width=\linewidth]{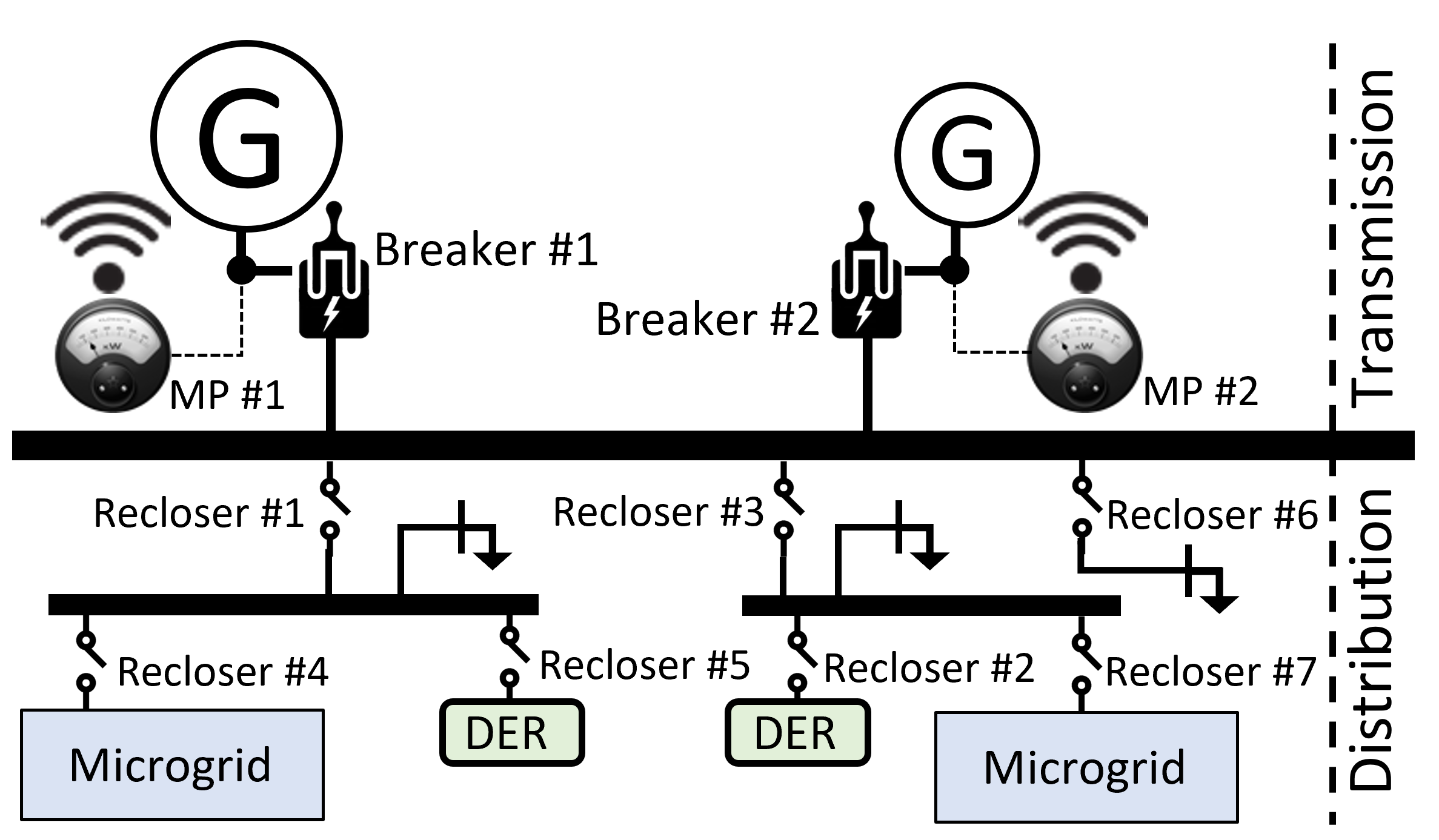}
    
    \caption{Simplified integrated T\&D model single line diagram.}
    \label{fig:sld}
    \vspace{-1mm}
    
\end{figure}

Fig. \ref{fig:sld} illustrates an architectural diagram of the integrated T\&D system under test. It is important to note that Fig. \ref{fig:sld} does not reflect the `actual' grid interconnections and topology; for confidentiality and security reasons, the integrated T\&D system model topology cannot be disclosed. Similarly, the nameplate capacities of generators, number of connected residential/commercial loads, microgrid characteristics, power flows, etc. are not provided since they would expose sensitive information regarding the actual power system architecture. Thus, we have sanitized the analysis results delineating critical information without open sourcing intelligence for the electric power critical infrastructure, which, if maliciously exploited, could endanger the operation of the power system. 
Equally important to the power system topology information is the location and control of  switching devices. We have underlined the importance of breakers and recloser switches in sectionalizing parts of the grid during adverse events impeding their spread system-wide; however, adversaries can leverage these mechanisms to compromise the system operation, leaving parts of it without power. For our analysis, we  mainly focus on two subcircuits on the distribution level since they arise as more prominent targets for adversaries compared to transmission systems which are typically better protected and monitored.

\begin{figure}[t!]
    \centering
    \includegraphics[width=\linewidth]{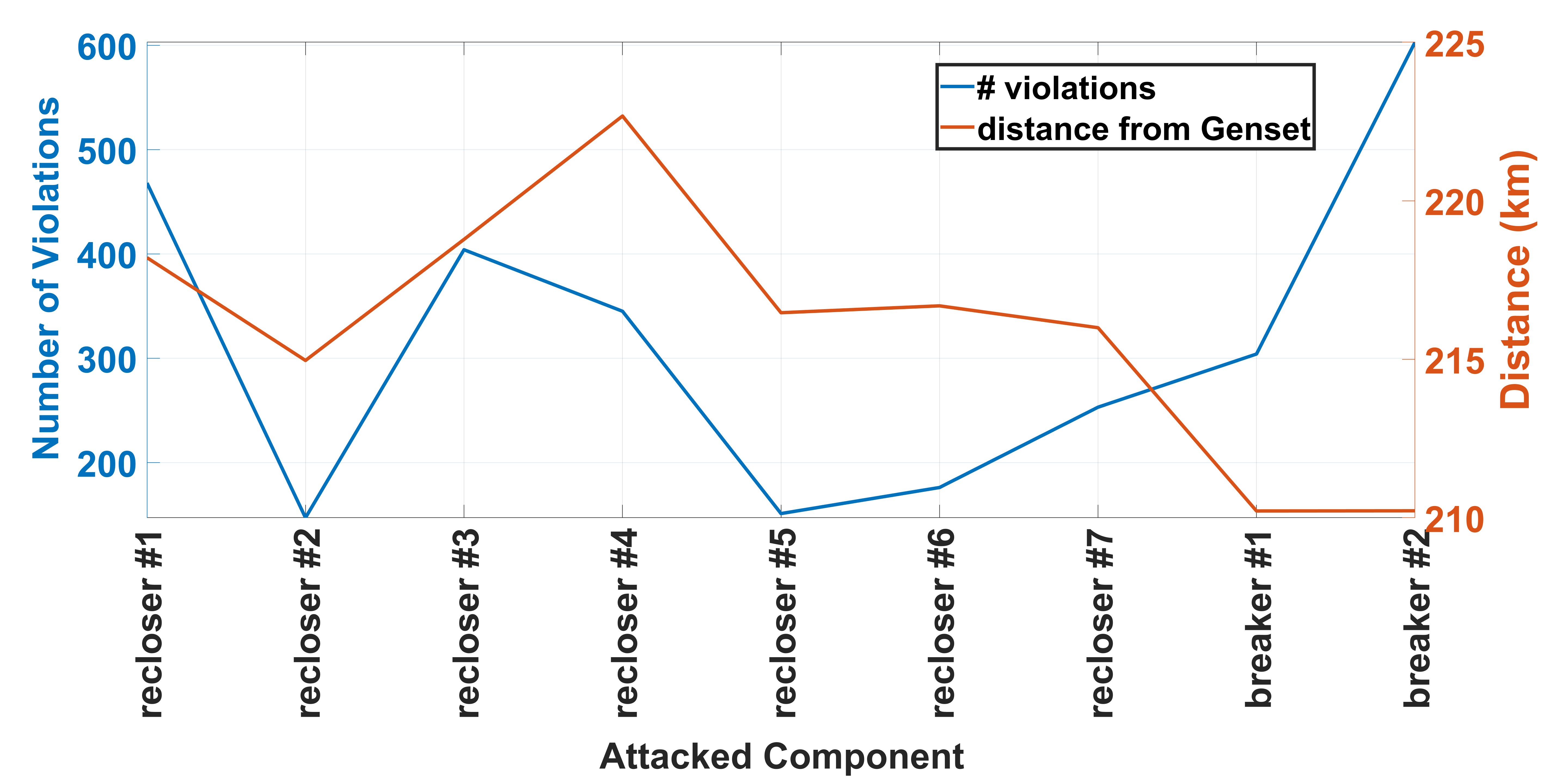}
    
    \caption{Switch and circuit breaker violations.}
    \label{fig:breaker}
    \vspace{-1mm}
    
\end{figure}

By performing power flow analysis and through maliciously modifying the behavior of SCADA controlled switches and breakers, we generate violation reports demonstrating the degree of impact introduced by such adversarial actions on the power system. The aforementioned violation reports (outlined in Fig. \ref{fig:breaker}) illustrate the most critical points for the system. Hence, their security should be prioritized since they would be the most favorable targets for adversaries aiming to maximize the inflicted damage. In Fig. \ref{fig:breaker}, we present the number of violations which occur in the integrated T\&D system model once any of the components (on the horizontal axis) gets compromised. Furthermore, the geographical distance between the attacked device and the generation facility is indicated. We observe that there is correlation between the device proximity to the generation point and the number of violations. If a device gets compromised, the successive devices on the same path will also be affected. Circuit breaker \#2 arises as the most vulnerable device (also closer to the generation point), followed by reclosers \#1 and \#3. 

Furthermore, although -- according to Fig. \ref{fig:breaker} -- breakers \#1 and \#2 are located at similar distances from their gensets, the number of their corresponding violations is significantly larger. This observation dictates that breaker \#2 is supporting more loads (out of which some might be critical loads, thus ``non-shedable''). The discrepancy between the reported violations implies that in the event of a breaker \#2 trip, the loads that do not belong to DER-enabled microgrids  (i.e., able to operate autonomously and meet the power demand) will have to be supported from neighbouring resources. Altering power dispatch to accommodate for such unexpected power demand leads to uneconomical grid operation, and hence, more violations. For our case study, the number of violations serves as an ``performance indicator''. In other words, more violations outline worse grid operation conditions.

\subsection{Interruption of system-critical operations: coordinated monitoring and switching attacks}
Situational awareness is essential in order to preserve power system reliability, stability, and mitigate the impact of adverse events such as blackouts and equipment failures. AMI and MPs enhance the observability of the power system states (e.g., voltage and current magnitude/angle, frequency, power, etc.) by providing regular updates to system operators. In this scenario, we consider time-delay attacks in which control commands to switches are delayed due to the lack of synchronization between system operation and MPs.

\begin{figure}[t]
    \centering
    \includegraphics[width=\linewidth]{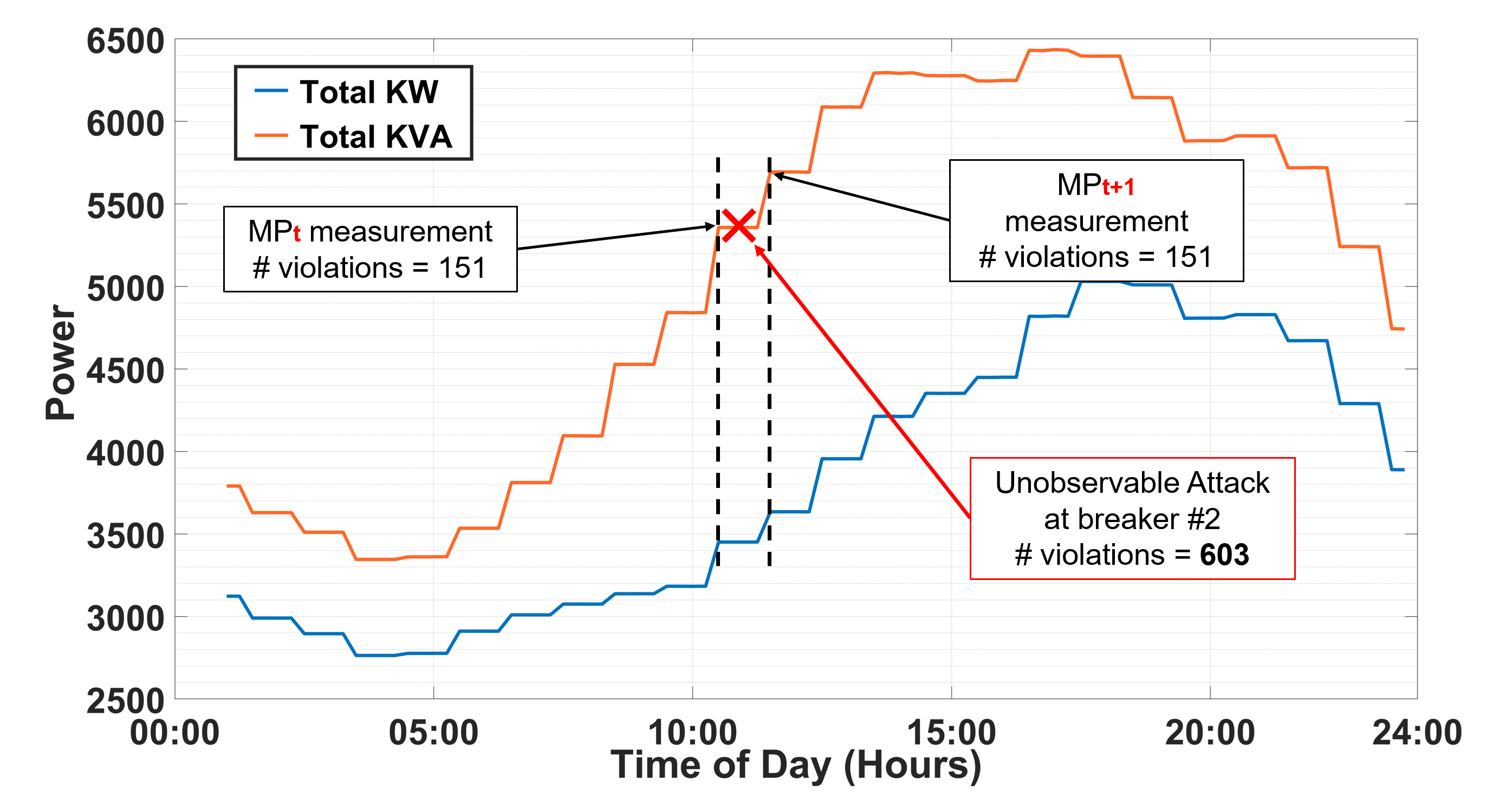}
    
    \caption{Monitor point (MP) frequency update granularity.}
    \label{fig:monitor}
    \vspace{-1mm}
    
\end{figure}

For our case study, the MPs are assumed to communicate with the control center (system operator side) at regular intervals, typically in the range of 10-15 minutes, thus short transient events can pass unnoticed if properly timed between system MP sampling intervals. The knowledge of the most critical component, i.e., circuit breaker \#2 per the previous attack study, can lead attackers to stealthily compromise the system operation. The aforementioned device can cause 603 violations. If the attack is properly synchronized, i.e., it occurs anywhere in the 15 minute window (Fig. \ref{fig:monitor}), detecting it becomes challenging. Thus, system operators can be oblivious to such severe events. Notably, the switching of the breaker is not noticed by the MP, and the reported values before switching the device remain unaltered, although we introduced a disconnect event between the MP sampling points. Similar attacks can be performed on different locations with varying system-wide impacts. In our implementation, we selected the most critical switching device, i.e., breaker \#2  to emphasize the corresponding effects.

\underline{Security Discussion:}
The goal of our analysis is to demonstrate that simulation-aided risk assessment is a useful tool for evaluating the condition of system operations. The proposed framework can be employed to identify power system related vulnerabilities, examine the effectiveness of mitigation strategies against such vulnerabilities, and design redundancy measures to overcome cyberattacks or unexpected failures. Our methodology serves as a proof-of-concept in this direction, and as such, the violation terminology can assume different definitions depending on the type of analysis being conducted and the nature of the system being investigated. For example, if we are simulating an attack scenario on switch and breaker controls, the number of violations could indicate the number of compromised/offline devices, loads being shed, overloaded lines, sectionalized power equipment, etc. On the other hand, if we are conducting a data integrity- or injection attack- type of study where adversaries aim to falsify the locational marginal price (LMP) mechanism, violations could be triggered if the electricity price surpasses a certain threshold yielding uneconomical grid operation. Regardless of the case study and the investigation-specific characteristics, the synergistic application of system simulations and risk assessment is beneficial in designing secure and resilient power systems.
\section{Conclusions} \label{s:conclusions}

In this paper, we present a cybersecurity analysis encompassing risk and impact assessment of power grids with DERs. Three different attack classes are discussed including data integrity attacks, alongside attacks aiming to block and and interrupt critical grid functions. Furthermore, we demonstrate how simulation-aware risk assessment analyses are critical for identifying vulnerable grid components. The security enhancement of such components could lead to more resilient power systems against cyberattacks. An integrated T\&D system model is used for the attack simulations and impact evaluations, while the OCTAVE Allegro methodology is utilized for the risk assessment process.

\section*{Disclaimer}

This paper was prepared as an account of work sponsored by an agency of the United States Government. Neither the United States Government nor any agency thereof, nor any of their employees, makes any warranty, express or implied, or assumes any legal liability or responsibility for the accuracy, completeness, or usefulness of any information, apparatus, product, or process disclosed, or represents that its use would not infringe privately owned rights. Reference herein to any specific commercial product, process, or service by trade name, trademark, manufacturer, or otherwise does not necessarily constitute or imply its endorsement, recommendation, or favoring by the United States Government or any agency thereof. The views and opinions of authors expressed herein do not necessarily state or reflect those of the United States Government or any agency thereof.

\bibliographystyle{IEEEtran}

{\bibliography{biblio}}

\begin{thebibliography}{10}
\providecommand{\url}[1]{#1}
\csname url@samestyle\endcsname
\providecommand{\newblock}{\relax}
\providecommand{\bibinfo}[2]{#2}
\providecommand{\BIBentrySTDinterwordspacing}{\spaceskip=0pt\relax}
\providecommand{\BIBentryALTinterwordstretchfactor}{4}
\providecommand{\BIBentryALTinterwordspacing}{\spaceskip=\fontdimen2\font plus
\BIBentryALTinterwordstretchfactor\fontdimen3\font minus
  \fontdimen4\font\relax}
\providecommand{\BIBforeignlanguage}[2]{{%
\expandafter\ifx\csname l@#1\endcsname\relax
\typeout{** WARNING: IEEEtran.bst: No hyphenation pattern has been}%
\typeout{** loaded for the language `#1'. Using the pattern for}%
\typeout{** the default language instead.}%
\else
\language=\csname l@#1\endcsname
\fi
#2}}
\providecommand{\BIBdecl}{\relax}
\BIBdecl

\bibitem{muyeen2017communication}
S.~Muyeen and S.~Rahman, \emph{Communication, control and security challenges
  for the smart grid}.\hskip 1em plus 0.5em minus 0.4em\relax The Institution
  of Engineering and Technology, 2017.

\bibitem{konstantinou2021secure}
C.~{Konstantinou}, ``Towards a secure and resilient all-renewable energy grid
  for smart cities,'' \emph{IEEE Consumer Electronics Magazine}, 2021.

\bibitem{jain2016three}
H.~Jain, A.~Parchure, R.~P. Broadwater, M.~Dilek, and J.~Woyak, ``Three-phase
  dynamic simulation of power systems using combined transmission and
  distribution system models,'' \emph{IEEE Transactions on Power Systems},
  vol.~31, no.~6, pp. 4517--4524, 2016.

\bibitem{tbaileh2017graph}
A.~Tbaileh, H.~Jain, R.~Broadwater, J.~Cordova, R.~Arghandeh, and M.~Dilek,
  ``Graph trace analysis: An object-oriented power flow, verifications and
  comparisons,'' \emph{Electric Power Systems Research}, vol. 147, pp.
  145--153, 2017.

\bibitem{bhatti2020analyzing}
B.~A. Bhatti, R.~Broadwater, and M.~Dilek, ``Analyzing impact of distributed pv
  generation on integrated transmission \& distribution system voltage
  stability—a graph trace analysis based approach,'' \emph{Energies},
  vol.~13, no.~17, p. 4526, 2020.

\bibitem{kuruvila2020hardware}
A.~Peedikayil~Kuruvila, I.~Zografopoulos, K.~Basu, and C.~Konstantinou,
  ``Hardware-assisted detection of firmware attacks in inverter-based
  cyberphysical microgrids,'' \emph{arXiv preprint arXiv:2009.07691}, 2020.

\bibitem{qi2016cybersecurity}
J.~Qi, A.~Hahn, X.~Lu, J.~Wang, and C.-C. Liu, ``Cybersecurity for distributed
  energy resources and smart inverters,'' \emph{IET Cyber-Physical Systems:
  Theory \& Applications}, vol.~1, no.~1, pp. 28--39, 2016.

\bibitem{konstantinou2015impact}
C.~Konstantinou and M.~Maniatakos, ``Impact of firmware modification attacks on
  power systems field devices,'' in \emph{Int'l Conference on Smart Grid
  Communications (SmartGridComm)}.\hskip 1em plus 0.5em minus 0.4em\relax IEEE,
  2015, pp. 283--288.

\bibitem{glenn2016cyber}
C.~Glenn, D.~Sterbentz, and A.~Wright, ``Cyber threat and vulnerability
  analysis of the us electric sector,'' Idaho National Lab.(INL), Idaho Falls,
  ID (United States), Tech. Rep., 2016.

\bibitem{johnson2017roadmap}
J.~Johnson, ``Roadmap for photovoltaic cyber security,'' \emph{Sandia National
  Laboratories}, 2017.

\bibitem{zografopoulos2020derauth}
I.~Zografopoulos and C.~Konstantinou, ``{DERauth: a battery-based
  authentication scheme for distributed energy resources},'' in \emph{IEEE
  Computer Society Annual Symposium on VLSI (ISVLSI)}, 2020, pp. 560--567.

\bibitem{zografopoulos2020harness}
I.~Zografopoulos, J.~Ospina, and C.~Konstantinou, ``{Special Session: Harness
  the Power of DERs for Secure Communications in Electric Energy Systems},'' in
  \emph{2020 IEEE 38th International Conference on Computer Design
  (ICCD)}.\hskip 1em plus 0.5em minus 0.4em\relax IEEE, 2020, pp. 49--52.

\bibitem{crownmitre}
{The MITRE Corporation}, ``{Crown Jewels Analysis},'' 2019.

\bibitem{9308900}
J.~{Ospina}, X.~{Liu}, C.~{Konstantinou}, and Y.~{Dvorkin}, ``On the
  feasibility of load-changing attacks in power systems during the covid-19
  pandemic,'' \emph{IEEE Access}, vol.~9, pp. 2545--2563, 2021.

\bibitem{jin2011event}
D.~Jin, D.~M. Nicol, and G.~Yan, ``An event buffer flooding attack in dnp3
  controlled scada systems,'' in \emph{Proceedings of the 2011 Winter
  Simulation Conference (WSC)}.\hskip 1em plus 0.5em minus 0.4em\relax IEEE,
  2011, pp. 2614--2626.

\bibitem{alcaraz2015critical}
C.~Alcaraz and S.~Zeadally, ``Critical infrastructure protection: Requirements
  and challenges for the 21st century,'' \emph{International journal of
  critical infrastructure protection}, vol.~8, pp. 53--66, 2015.

\bibitem{NISTSP}
\BIBentryALTinterwordspacing
{National Institute of Standards and Technology (NIST)}, ``{SP 800-82 Rev. 2 --
  Guide to Industrial Control Systems (ICS) Security},'' 2015. [Online].
  Available:
  \url{https://csrc.nist.gov/publications/detail/sp/800-82/rev-2/final}
\BIBentrySTDinterwordspacing

\bibitem{NISTIR}
\BIBentryALTinterwordspacing
{National Institute of Standards and Technology~(NIST)}, ``{NISTIR 7628 Rev. 1
  -- Guidelines for Smart Grid Cybersecurity},'' 2014. [Online]. Available:
  \url{https://csrc.nist.gov/publications/detail/nistir/7628/rev-1/final}
\BIBentrySTDinterwordspacing

\bibitem{zografopoulos2021cyberphysical}
I.~{Zografopoulos}, J.~{Ospina}, X.~{Liu}, and C.~{Konstantinou},
  ``{Cyber-Physical Energy Systems Security: Threat Modeling, Risk Assessment,
  Resources, Metrics, and Case Studies},'' \emph{IEEE Access}, vol.~9, pp.
  29\,775 -- 29\,818, 2021.

\bibitem{ospina2020demo}
J.~Ospina, I.~Zografopoulos, X.~Liu, and C.~Konstantinou, ``Demo: Trustworthy
  cyberphysical energy systems: Time-delay attacks in a real-time co-simulation
  environment,'' in \emph{Proceedings of the 2020 Joint Workshop on CPS\&IoT
  Security and Privacy}.\hskip 1em plus 0.5em minus 0.4em\relax ACM, 2020,
  p.~69.

\bibitem{8743447}
O.~M. {Anubi} and C.~{Konstantinou}, ``Enhanced resilient state estimation
  using data-driven auxiliary models,'' \emph{IEEE Transactions on Industrial
  Informatics}, vol.~16, no.~1, pp. 639--647, 2020.

\bibitem{hedman2011review}
K.~W. Hedman, S.~S. Oren, and R.~P. O'Neill, ``A review of transmission
  switching and network topology optimization,'' in \emph{2011 IEEE power and
  energy society general meeting}.\hskip 1em plus 0.5em minus 0.4em\relax IEEE,
  2011, pp. 1--7.

\bibitem{muller1989line}
N.~M{\"u}ller and V.~Quintana, ``Line and shunt switching to alleviate
  overloads and voltage violations in power networks,'' in \emph{IEE
  Proceedings C -- Generation, Transmission and Distribution}, vol. 136,
  no.~4.\hskip 1em plus 0.5em minus 0.4em\relax IET, 1989, pp. 246--253.

\end{thebibliography}

\end{document}